\documentstyle[12pt]{article}

\def\lsim{\raise0.3ex\hbox{$\;<$\kern-0.75em\raise-1.1ex\hbox{$\sim\;$}}}
\def\gsim{\raise0.3ex\hbox{$\;>$\kern-0.75em\raise-1.1ex\hbox{$\sim\;$}}}
\def\JP{\hbox{\sf J\kern-0.4em\raise0.1ex\hbox{P}}}
\def\JTP{\hbox{\sf J\kern-0.4em\raise0.1ex\hbox{P\kern-0.5em\hbox{T}}}}
\newcommand {\unit}[1]{\hspace{0.33em} {\rm #1}}
\newcommand {\ignore}[1]{}

\newcommand{\noi}{\noindent}
\newcommand{\bc}{\begin{center}}
\newcommand{\ec}{\end{center}}

\def\ifmath#1{\relax\ifmmode #1\else $#1$\fi}
\def\3quarter{{\textstyle{3 \over 4}}}

\def\vs{\vskip}

\def\ra{\rightarrow}

\def\lf{\leaders\hbox to 1em{\hss.\hss}\hfill}

\def\21{$SU(2) \ot U(1)$}
\def\O{\hbox{$\cal O$ }}

\def\etal{\hbox{\it et al., }}
\def\J.W.F.V{\hbox{J. W. F. Valle }}
\def\gau{\hbox{gauge }}

\def\sm{\hbox{standard model }}

\def\neu{\hbox{neutrino }}

\def\eq#1{{eq. (\ref{#1})}}

\def\VEV#1{\left\langle #1\right\rangle}

\def\lsim{\raise0.3ex\hbox{$\;<$\kern-0.75em\raise-1.1ex\hbox{$\sim\;$}}}
\def\gsim{\raise0.3ex\hbox{$\;>$\kern-0.75em\raise-1.1ex\hbox{$\sim\;$}}}
\def\bel{\begin{letter}}
\def\eel{\end{letter}}
\def\beq{\begin{equation}}
\def\eeq{\end{equation}}
\def\bef{\begin{figure}}
\def\eef{\end{figure}}
\def\bet{\begin{table}}
\def\eet{\end{table}}
\def\bea{\begin{eqnarray}}
\def\ba{\begin{array}}
\def\ea{\end{array}}
\def\bi{\begin{itemize}}
\def\ei{\end{itemize}}
\def\ben{\begin{enumerate}}
\def\een{\end{enumerate}}
\def\ra{\rightarrow}
\def\ot{\otimes}
\def\eea{\end{eqnarray}}

\def\ib#1#2#3{           {\it ibid. }{\bf #1} (19#2) #3}

\def\np#1#2#3{           {\it Nucl. Phys. }{\bf #1} (19#2) #3}
\def\pl#1#2#3{           {\it Phys. Lett. }{\bf #1} (19#2) #3}
\def\pr#1#2#3{           {\it Phys. Rev. }{\bf #1} (19#2) #3}

\def\n.c.#1#2#3{         {\it Nuovo Cim. }{\bf #1} (19#2) #3}
\def\r.n.c.#1#2#3{       {\it Riv. del Nuovo Cim. }{\bf #1} (19#2) #3}
\def\sjnp#1#2#3{         {\it Sov. J. Nucl. Phys. }{\bf #1} (19#2) #3}

\relax
\def\caption{\refstepcounter\@captype \@dblarg{\@caption\@captype}}

\long\def\@caption#1[#2]#3{\addcontentsline{\csname
  ext@#1\endcsname}{#1}{\protect\numberline{\csname
  the#1\endcsname}{\ignorespaces #2}}\par
  \begingroup
    \@parboxrestore
    \small                                      %---
    \@makecaption{\csname fnum@#1\endcsname}{\ignorespaces #3}\par
  \endgroup}

\newlength{\anchocaption}                       %---

\long\def\@makecaption#1#2{
   \vskip 10pt
   \setbox\@tempboxa\hbox{#1: #2}
   \setlength{\anchocaption}{\hsize}            %---
   \addtolength{\anchocaption}{-2\leftmargini}  %---
   \ifdim \wd\@tempboxa >\anchocaption   % IF longer than one line:     %---
        \begin{list}{}{ \setlength{\leftmargin}{\leftmargini}           %---
                        \setlength{\rightmargin}{\leftmargini}          %---
                        \setlength{\labelsep}{\leftmargini}             %---
                        \setlength{\labelwidth}{0pt} }                  %---
           \item[] #1: #2                                               %---
        \end{list}                                                      %---
     \else                        %   ELSE  center.
        \begin{center} #1: #2 \end{center}                              %---
   \fi}

\def\chapi{\par
 \setcounter{chapter}{0}
 \setcounter{section}{0}
 \def\@chapapp{Chapter}
 \def\thechapter{\arabic{chapter}}}

\def\chap{\par
 \setcounter{chapter}{0}
 \setcounter{section}{0}
 \def\@chapapp{Cap\'{\i}tulo}
 \def\thechapter{\arabic{chapter}}}

\def\appendix{\par
 \setcounter{chapter}{0}
 \setcounter{section}{0}
 \def\@chapapp{Ap\'{e}ndice}
 \def\thechapter{\Alph{chapter}}}

\def\appendixi{\par
 \setcounter{chapter}{0}
 \setcounter{section}{0}
 \def\@chapapp{Appendix}
 \def\thechapter{\Alph{chapter}}}

\begin{document}
%\tableofcontents
\begin{titlepage}
%\today
\begin{center}
\hfill FTUV/92-60\\
\hfill IFIC/92-60\\
\vskip 0.3cm
{\Large \bf Massive Neutrinos and Electroweak Baryogenesis}\\
\vspace{2cm}
\vskip .3cm
{\Large J.T. Peltoniemi }
\footnote{E-mail peltonie at vm.ci.uv.es or 16444::peltonie}\\
and\\
{\Large Jos\'{e} W. F. Valle}
\footnote{E-mail valle at vm.ci.uv.es or 16444::valle}\\
\vs .2cm
\noi
{\it Instituto de F\'{\i}sica Corpuscular - C.S.I.C.\\
Departament de F\'{\i}sica Te\`orica, Universitat de Val\`encia\\
46100 Burjassot, Val\`encia, SPAIN}\\
\vs .5cm
{\large \bf Abstract}
\end{center}
\vspace{2cm}
\begin{quotation}
We show that certain models for neutrino masses
that lead to the MSW explanation of the solar \neu
data and/or the hot dark matter component in the universe
can also naturally allow for the successful generation of the
cosmological baryon asymmetry at the electroweak scale.
This follows either because the experimental Higgs boson
mass limit is weakened due to the smaller production
rate and the presence of sizeable invisible decays
involving majoron emission,
or due to the effect of additional electrically charged
scalar bosons which substantially modify the condition
for succesful baryogenesis.
\end{quotation}
\end{titlepage}

\section{Introduction}

In order to generate the cosmological baryon asymmetry
one needs sufficient CP-violation, B violation and
out of equilibrium conditions \cite{Sakharov67}.
These requirements may be satisfied in
grand-unified theories.
The observation that nonperturbative B violating
electroweak effects \cite{tHooft76} can be substantial
at high temperatures, as expected to hold in the early
universe  opens the possibility of baryogenesis
at the weak scale \cite{Kuz}.
On the other hand, even if the baryon asymmetry is due to
physics at superhigh scales, it may be washed out during
the electroweak phase transitions, as stressed
by Shaposhnikov and others \cite{Kuz,Shaposhnikov88,Linde,Dineetal92a}.
Although the exact details are still the subject of
controversy, there is now a consensus that, for
reasonable choices of the
Higgs boson mass, the transition can be first
order, as required for electroweak baryogenesis.
However, although the Standard Model may satisfy
the formal requirements for baryogenesis, is unlikely
to provide, on its own, a successful model.

In order to prevent that the baryon asymmetry
that might be generated at the weak scale
(or at the unification scale)
be washed out during the phase transition
one requires \cite{Dineetal92a}
\beq
\label{40a}
m_H \lsim 40 \rm{GeV}
\eeq
Unfortunately this upper bound on the Higgs
boson mass required for successful baryogenesis
is in conflict with the recent combined
lower limit obtained
from the four LEP experiments \cite{LEP1}, i.e.
\beq
\label{60}
m_H \geq 60 \rm{GeV}
\eeq

The existence of new Higgs fields provides
new mechanisms which resolve this discrepancy
and satisfy the criteria required for weak scale
baryogenesis. Some schemes of this
type have already been suggested in the
literature \cite{AndersonHall92,2Higgs}.

In this Letter we stress that several
models which have been proposed for different
reasons predict an extended Higgs sector.
We focus on models %%are motivated by the fact that they
that can naturally account for the small neutrino
masses \cite{fae} suggested by the MSW explanation \cite{MSW}
of present solar neutrino data \cite{Davis_SAGE_GALLEX1_KAMII}
and/or by cosmological observations suggesting the need
for a hot dark matter component in the universe \cite{cobe}.
In this class of models we identify two ways
whereby the conflict between weak scale
baryogenesis and LEP data can be avoided:
\ben
\item
There are new electrically charged scalar bosons
whose effects can substantially change the condition
in \eq{40a}, allowing much larger masses.
\item
There are sizeable invisible Higgs decays
involving singlet majoron emission. In this
case the bound in \eq{60} does not hold
since the rates for Higgs boson production
are lower than the standard model prediction,
as noted also in ref. \cite{EnqvistKainulainenVilja92}.
\een
We provide a simple model example
and discuss some of its features.

\section{Prototype Model}

For definiteness we focus on a simple
representative \cite{Babu88} of a class of models which
can naturally generate neutrino masses in
the range $10^{-3}$ eV where the MSW effect
provides an explanation of the solar \neu data.
The model contains, in addition to the Standard
Model particles, one singly charged singlet scalar $\eta^-$
and one doubly charged singlet scalar $\chi^{--}$.
The scalar boson lagrangean is given by
\begin{eqnarray} \label{}
V&=&
-\frac{1}{2}\mu_\phi^2 \phi ^\dagger \phi + \frac{1}{4}\lambda_\phi
(\phi ^\dagger \phi)^2
+\frac{1}{2}\mu_\chi^2 \chi ^\dagger \chi +
\frac{1}{4}\lambda_\chi (\chi ^\dagger \chi)^2
+\frac{1}{2}\mu_\eta^2 \eta ^\dagger \eta + \frac{1}{4}\lambda_\eta
(\eta ^\dagger \eta)^2 \nonumber\\&&
+\frac{1}{2}\epsilon_{\phi \chi } \phi ^\dagger \phi \chi ^\dagger
\chi
+\frac{1}{2}\epsilon_{\phi \eta } \phi ^\dagger \phi \eta ^\dagger \eta
+\frac{1}{2}\epsilon_{\eta \chi } \eta^\dagger \eta \chi^\dagger \chi
+ \mu_{cubic} \chi^{++} \eta^- \eta^- + H.c.
\end{eqnarray}
where $\epsilon$ and $\lambda$ are dimensionless coupling
constants and $\mu$'s are mass parameters. The Higgs boson
vacuum expectation value (at zero temperature) responsible
for electroweak breaking is given by
\begin{eqnarray}
\label{}
\VEV{\phi} & = & \frac{\mu_\phi }{\sqrt{\lambda _\phi }}
\end{eqnarray}
while the $\chi$ and $\eta$ mass terms are chosen positive
in such a way that the electrically charged Higgs bosons
do not acquire vacuum expectation values, as required.

With these scalars and the following lepton Yukawa
interactions
\beq
{\cal L} = -\frac{\sqrt2 m_i}{v}{\bar{\ell}}_i \phi e_{Ri} +
f_{ij} \ell_i^T C i \tau_2 \ell_j \eta^+ +
h_{ij} e_{Ri}^T C e_{Rj} \chi^{++} + H.c.
\label{yuk}
\eeq
($h$ and $f$ are symmetric and anti-symmetric
coupling matrices, respectively) the lepton masses
are generated when the \21 \gau symmetry is
broken by $\VEV{\phi}$. The first term gives
the charged lepton masses $m_i$ at the tree
level, while neutrinos acquire masses radiatively,
at the two loop level, by the diagram in Fig. 1.
For natural choices of parameters,
consistent with all present observations,
e.g. $f_{e\tau},  f_{\mu\tau}, h_{\tau\tau}
\simeq 0.05$, $\mu_{cubic} \simeq 1$ GeV,
and charged Higgs  boson masses of about 200 GeV,
these \neu masses are in the $10^{-3}$ eV range.

We will now study the consequences of the extended
Higgs sector for the electroweak phase transition.
At high temperatures, the effective potential of the
standard Higgs field
responsible for the electroweak phase transition can
be written as \cite{DolanJackiw,AndersonHall92} %{Dineetal92a,Dineetal92b}
\beq
\label{}
V_\phi (T) = D ( T^2 - T_0^2 ) \phi^2
+ ET \phi^3  + \frac{\lambda_T}{4} \phi^4 .
\eeq
The effect of the additional Higgs bosons appears
as modifications in the coefficients of the finite
temperature effective potential. In the following
we assume that $\mu_\alpha \ll \epsilon_{\alpha \phi} \VEV{\phi}$,
$ \alpha =\eta, \phi $ in which case
the effect of the new scalars is maximal.
The coefficient of the quadratic term is given by
\begin{eqnarray}
\label{}
D &=& \frac{6 M_W^2 + 3M_Z^2 + 6m_t^2 + 2M_\eta^2 + 2M_\chi^2} {24
\VEV{\phi}^2}
\end{eqnarray}
where
\begin{eqnarray}
\label{}
T_0 &=& \sqrt{\frac{\mu_\phi^2 - 8 B \VEV{\phi}^2}{4D}}\\
B &=& \frac{6M_W^4+3M_Z^4-12m_t^4+2M_\eta^4+2M_\chi^4 }{64\pi^2
\VEV{\phi}^4 }
\end{eqnarray}
The coefficient of the cubic term, taking into account
higher order radiative corrections involving \gau boson
effects \cite{Dineetal92b}, is given by
\beq
\label{}
E = \frac{1}{6 \pi \VEV{\phi}^3}(2 M_W^3 +  M_Z^3 +
M_\eta^3 + M_\chi^3)
\eeq
and that of the quartic term is
\begin{eqnarray}
\label{}
\lambda _T &=& \lambda _\phi -\frac{1}{16\pi ^2 \VEV{\phi}^4}
 \left( 6 M_W^4 \ln\frac{M_W^2}{a_b T^2}
+ 3 M_Z^4 \ln\frac{M_Z^2}{a_b T^2} \nonumber \right. \\ && \mbox{ }
\left. \mbox{ }
+ 2 M_\eta^4 \ln\frac{M_\eta^2}{a_b T^2}
+ 2 M_\chi^4 \ln\frac{M_\chi^2}{a_b T^2}
- 12 m_t^4 \ln\frac{m_t^2}{a_f T^2} \right)
\end{eqnarray}
where $\ln a_b\approx 3.91$ and $\ln a_f \approx 1.14$, using the
renormalisation conditions of \cite{Dineetal92b}.
%Above was assumed that masses were generated solely by $\phi$. If the term
%$\mu_\eta$ gives a substantial contribution, one should replace $M_\eta^2$ by
%$M_\eta^2-\mu_\eta^2$.

The washout of the baryon asymmetry generated by
sphaleron processes would occur during the phase
transition, as the Higgs field tunnels from the false
vacuum to the true vacuum. The degree of washout depends
on the tunneling conditions, which in turn depend on the
width and height of the potential barrier.
In order to prevent this washout one requires
that $\phi_C/T_C \gsim 1$ \cite{Shaposhnikov88}
at the critical moment when the phase transition
starts. The results from numerical calculations
\cite{Dineetal92b} yield
\beq
\label{}
\phi_C \simeq \frac{2.4 E}{\lambda_{T_C}} T_C
\eeq
Adopting this value we
%find that the washout
%requirement is then $E/\lambda \gsim 0.44$,
%which gives
obtain the following constraint for the Higgs boson mass
\beq
\label{40b}
M_H \lsim 40 \unit{GeV}
\sqrt{1 + \left(\frac{M_\eta}{110 \unit{GeV}}\right)^3 +
          \left(\frac{M_\chi }{110 \unit{GeV}} \right)^3
        - 0.4   \left(\frac{M_\chi }{110 \unit{GeV}} \right)^4
        - 0.4   \left(\frac{M_\eta }{110 \unit{GeV}} \right)^4 },
\eeq
valid for singlet scalar masses up to a couple of hundreds GeV.
This bound is in agreement with the experimental limit
of \eq{60} if either of the scalars is heavier than
about 170 GeV, or if they are equally heavy, both
heavier than about 110 GeV, as illustrated in Fig. 2.
This condition is by no means unnatural, and is
in agreement with all experimental restrictions from
standard model Higgs boson searches. Likewise, it provides
an explanation of the solar neutrino data, as seen above.

\section{Introducing the Majoron}

Next we consider the variant of the above model
which we obtain by adding a neutral singlet $\sigma$
carrying two units of lepton number defined by the
$\sigma \chi \eta \eta$ coupling.
We postulate that this lepton number symmetry is
broken spontaneously by a nonzero vacuum expectation
value $\VEV{\sigma}$ of the $\sigma$ field. In this
case there is no cubic term $\mu_{cubic}$, this
now being replaced by the quartic
$\sigma \chi \eta \eta$ coupling.
For simplicity we take $\VEV{\sigma}$ to be small,
say \O (1GeV) as the scale for the breaking of
the lepton number symmetry. Clearly this fits
nicely with the evaluation of the \neu masses given
above.
To the extent that $\VEV{\sigma} \ll \VEV{\phi}$
we may consider the two phase transitions separately.
During the electroweak phase transition the $\sigma$ field
is still in the unbroken phase, and it behaves
as the charged fields of the previous example.
This way we avoid the complications of ref.
\cite{EnqvistKainulainenVilja92}, and apply
directly the results of Anderson and Hall
\cite{AndersonHall92}.

Note also that if $\VEV{\sigma}$ is small the
${\sigma}$ field will not give a significant
contribution to the Higgs potential
and thus the charged fields still play the
roles they had previously. Hence, adding
the majoron does not affect the $upper$ bound
in \eq{40b}. However, it crucially affects the
Higgs boson $lower$ mass limit derived from
the LEP experiments \cite{HJJ,JoshipuraValle92}.

The addition of the neutral singlet $\sigma$
provides two additional electrically neutral
degrees of freedom, an extra CP even Higgs boson
and a CP odd scalar boson, the majoron $J$.
The latter remains massless as a result of
the spontaneous nature of the lepton number
breaking. Its existence implies additional
decay modes for the two neutral CP even mass
eigenstate scalar bosons $H_1$ and
$H_2$. They both can decay by majoron emission,
either directly, such as $H_1 \ra JJ$ and
$H_2 \ra JJ$ or indirectly, via $H_2 \ra H_1 H_1$.
The coupling of $H_i$ to the majoron $J$ can
be written as
\begin{equation}
\label{J1}
{\cal L}_J=
\frac{(\sqrt 2 G_F)^{1/2}}{2}
\frac{\VEV{\phi}}{\VEV{\sigma}}
[M_2^2 \cos \theta H_2-M_1^2 \sin \theta H_1]J^2
\end{equation}
where the mixing angle $\theta$ and the Higgs masses $M_i^2$
are determined by the parameters of the Higgs potential.
The  masses $M_{1,2}^2$, the mixing angle $\theta$, and
the ratio $\frac{\VEV{\phi}}{\VEV{\sigma}}$ can be taken
as independent parameters in terms of which all couplings
can be fixed.
One sees from \eq{J1} that, as long as the scale at which
the lepton number breaking is sufficiently low these {\sl dark} Higgs
decay modes will be substantial, suppressing the
Higgs branching ratios into $b\bar{b}$.

On the other hand the production of the two
CP even Higgs bosons through the Bjorken process
$e^+ e^- \ra Z H_i$ will be smaller than predicted
in the \sm by factors $\sin^2\theta$ and $\cos^2\theta$.
For appropriately chosen values for the mixing angle $\theta$
the limit from \eq{60} can be completely avoided. A model
of this type was discussed in ref. \cite{HJJ}.

Both effects discussed above, i.e. smaller
production rates and the presence of sizeable
invisible Higgs decays will weaken the
experimental limit for the Higgs mass
so that \eq{60} is not applicable. Thus
we conclude that the presence of the
majoron alone would be enough to remove the
conflict with electroweak baryogenesis.
However, we also have, in addition,
the effects of the charged scalar bosons.
As a result, to a good approximation
the upper bound in \eq{40b} is still
valid and replaces that of \eq{40a}.

\section{Discussion}

We have described a model that may be
useful in generating a baryon excess at the weak scale.
The Higgs potential already has the necessary out of
equilibrium phase transition, and the electroweak
interactions can violate baryon number. In addition
CP violation can be introduced both in the Higgs
potential as well as in the Yukawa interaction.
The model retains most of the simplicity of the
\21 Higgs sector, including the canonical value
$\rho = 1$ for the ratio of neutral to charged
current strengths. It can realize two ways
to avoid the conflict between weak scale
baryogenesis and LEP data, either as a result
of the existence of electrically charged
singlet scalar bosons whose effects
modify condition \eq{40a} to \eq{40b},
allowing much larger masses, or as a result
of the softening of the LEP bound due to
the existence of sizeable invisible Higgs
decays with majoron emission, as we have
described. From this point of view it will be
interesting to perform dedicated searches
for a heavy invisibly decaying Higgs boson.

The model we presented was just an illustration.
A large class of similar models motivated by \neu
physics have been built \cite{JoshipuraValle92}.
They may generate \neu masses as required by
the existence of a hot dark matter component
in the universe.
Models also exist that can reconcile these
dark matter neutrinos with the solar \neu
deficit \cite{DARK92}. Many of these schemes
involve radiative generation of \neu masses due
to the existence of new Higgs bosons. As a result
the effect described in section 2 is present
and the electroweak baryogenesis is viable.
In addition, in all these models one can have
the spontaneous violation of lepton number
symmetry at or below the electroweak scale
by an \21 singlet vacuum expectation value
$\VEV{\sigma} \lsim \O(1)$ TeV. This leads
to a substantial invisible Higgs decay
channel with majoron emission and to the
possibility of a smaller rate of production
in Z decays. This will soften the Higgs mass
bounds obtained from LEP, removing again the
conflict with electroweak baryogenesis.

%\vs 2cm
\vfill
We thank Daniele Tommasini for helpful discussions
and Fernando de Campos for help with plotting.
This work was supported by CICYT and by the Spanish
Ministry of Science.

\newpage
\section*{Figure Captions}
\noindent
{\bf Fig. 1. }\\
Diagram generating small Majorana neutrino masses.\\
{\bf Fig. 2.}\\
Region of electrically charged Higgs boson masses
in GeV consistent with electroweak baryogenesis and
standard model Higgs boson searches at LEP. These
masses also allow an explanation of the solar
neutrino data by the MSW effect.
\newpage

%\bibliographystyle{ansrt}
%\bibliography{biblio}

\begin{thebibliography}{10}%\input{bibin}

\bibitem{Sakharov67}
 A.~D. Sakharov,
\newblock {\em Pisma v ZhETF} {\bf 5},32 (1967).

\bibitem{tHooft76}
 G. t'Hooft,
\newblock {\em Phys.\ Rev.\ Lett.} {\bf 37},8 (1976).

\bibitem{Kuz}
 V.~A. Kuzmin,  V.~A. Rubakov, and  M.~E. Shaposhnikov,
\newblock {\em Phys.\ Lett.} {\bf 155B},36 (1985).

\bibitem{Shaposhnikov88}
M Shaposhnikov, XXIV Int. Conference on High Energy Physics,
p. 1549, ed R. Kotthaus and J K\"uhn, Springer, 1988.

\bibitem{Linde}
A. Linde, {\it Particle Physics and
Inflationary Cosmology} (Harwood Academic Publishers 1990)


\bibitem{Dineetal92a}
 M. Dine,  R.~L. Leigh,  P. Huet,  A. Linde, and  D. Linde,
\newblock {\em Phys.\ Lett.\ B} {\bf 283},319 (1992).
M.E. Carrington, \pr{D45}{92}{2933}

\bibitem{LEP1}
 J. Steinberger,
\newblock in {\em Electroweak Physics Beyond the Standard Model},  ed.\  J.
  W.~F. Valle and  J. Velasco (World Scientific Publ. Co., Singapore, 1992)
  p.~3.

\bibitem{2Higgs}
A. I.  Bocharev, S.V. Kuzmin and M.E. Shaposnikov,
\pl{B244}{90}{275}; \pr{D43}{91}{369};
N. Turok and J. Zadrozny, \np{B358}{91}{471};
B. Kastening, R.D. Peccei and X. Zhang, \pl{B266}{91}{413};
L. McLerran \etal\pl{B256}{91}{451}


\bibitem{AndersonHall92}
 G.~W. Anderson and  L.~J. Hall,
\newblock {\em Phys.\ Rev.\ D} {\bf 45},2685 (1992).


\bibitem{Davis_SAGE_GALLEX1_KAMII}
 J. R.~Davis et~al.,
\newblock in {\em Proceedings of the 21th International Cosmic Ray Conference,
  Vol. 12},  ed.\  R.~J. Protheroe (University of Adelaide Press, 1990) p. 293.
%\bibitem{SAGE}
 A. Abazov et~al.,
\newblock {\em Phys.\ Rev.\ Lett.} {\bf 67},3332 (1991).
%\bibitem{GALLEX1}
 P. Anselmann et~al.,
\newblock {\em Phys.\ Lett.} {\bf B285},376 (1992).
%\bibitem{KAMII}
 K. Hirata et~al.,
\newblock {\em Phys.\ Rev.\ Lett.} {\bf 65},1297 (1990)
\newblock ;\ib{65}{90}{1301} and \ib{66}{91}{9}.

\bibitem{fae}
 J. W.~F. Valle,
\newblock {\em Prog.\ Part.\ Nucl.\ Phys.} {\bf 26},91 (1991)
\newblock and references therein.

\bibitem{MSW}
M Mikheyev, A Smirnov, \sjnp{42}{86}{913};
L Wolfenstein, \pr {D17}{78}{2369};\ib{D20}{79}{2634}.

\bibitem{cobe}
 G.~F. Smoot et~al.,
\newblock {\em Astrophys. J. Lett.} {\bf 396},L1 (1992)


\bibitem{Babu88}
 K.~S. Babu,
\newblock {\em Phys. Lett. B} {\bf 203}, 132 (1988).

\bibitem{DolanJackiw}
L. Dolan and R. Jackiw, {\em Phys.\ Rev.\ D} 9, 3320 (1974)

\bibitem{Dineetal92b}
 M. Dine,  R.~L. Leigh,  P. Huet,  A. Linde, and  D. Linde,
\newblock {\em Phys.\ Rev.\ D} 46, 550 (1992).

\bibitem{EnqvistKainulainenVilja92}
 K. Enqvist,  K. Kainulainen, and  I. Vilja,
\newblock NORDITA preprint 92/10P (1992).

\bibitem{HJJ}
 J.~C. Romao,  F. de~Campos, and  J. W.~F. Valle,
\newblock {\em Phys.\ Lett.} {\bf B292}, 329 (1992).

\bibitem{JoshipuraValle92}
 A. Joshipura and  J. W.~F. Valle,
\newblock CERN preprint TH.6652 (1992).

\bibitem{DARK92}
 J. T. Peltoniemi, D. Tommasini, and  J. W.~F. Valle,
\newblock {\em CERN-TH.6624}, {\em Phys.\ Lett. B} (1993),
in press

\end{thebibliography}

\end{document}